\documentstyle[12pt,epsfig]{article}

{\end{list}}
\newcounter{enumct}

\begin{document}
 
\sloppy


\thispagestyle{empty}
\begin{flushright}
\large
DTP/98/64 \\ October 1998
\end{flushright}
\vspace{0.65cm}

\begin{center}
\LARGE
{\bf Parton Content of Polarized Photons:}

\vspace{0.1cm}
\hbox to\textwidth{\hss
{\bf Theoretical Status and Experimental Prospects}
\hss}

\vspace{1.2cm}
\Large
M.\ Stratmann\\
\vspace{0.5cm}
\large
Department of Physics, University of Durham,\\ 
\vspace{0.1cm}
Durham DH1 3LE, England\\
\vspace{2.4cm}
{\bf Abstract} 
\end{center}
\vspace*{0.5cm}

\noindent
The theoretical framework for a next-to-leading order QCD analysis
of the spin-dependent parton densities $\Delta f^{\gamma}(x,Q^2)$ of the
longitudinally polarized photon 
and of its structure function $g_1^{\gamma}(x,Q^2)$ is reviewed in some
detail. Special emphasis is given to the specific features of 
different factorization schemes. Two conceivable scenarios for the
polarized parton densities $\Delta f^{\gamma}(x,Q^2)$ are introduced,
which are suitable to study the sensitivity of future experiments
to the so far unmeasured $\Delta f^{\gamma}$.
The experimental prospects of determining $\Delta f^{\gamma}$
at a polarized collider mode of HERA or at a linear $e^+e^-$ collider
are outlined.
Finally, the $Q^2$-evolution of the parton content of 
linearly polarized photons, which shows
some remarkable differences, is briefly discussed.
\normalsize

\vspace{6.0cm}
\noindent
{\it Talk presented at the workshop on
`Photon Interactions and the Photon Structure', Lund, Sweden,
September 1998.}
\vfill

\setcounter{page}{0}
\newpage

\begin{center}
{\LARGE\bf Parton Content of Polarized Photons: }\\[4mm]
\hbox to\textwidth{\hss
{\LARGE\bf Theoretical Status and Experimental Prospects}\hss}
\vspace*{13mm}
{\Large Marco Stratmann} \\[3mm]
{\it Department of Physics, University of Durham, 
Durham DH1 3LE, England}\\[1mm]
{\it E-mail: Marco.Stratmann@durham.ac.uk}\\[20mm]

{\bf Abstract}\\[1mm]
\begin{minipage}[t]{140mm}
The theoretical framework for a next-to-leading order QCD analysis
of the spin-dependent parton densities $\Delta f^{\gamma}(x,Q^2)$ of the
longitudinally polarized photon 
and of its structure function $g_1^{\gamma}(x,Q^2)$ is reviewed in some
detail. Special emphasis is given to the specific features of 
different factorization schemes. Two conceivable scenarios for the
polarized parton densities $\Delta f^{\gamma}(x,Q^2)$ are introduced,
which are suitable to study the sensitivity of future experiments
to the so far unmeasured $\Delta f^{\gamma}$.
The experimental prospects of determining $\Delta f^{\gamma}$
at a polarized collider mode of HERA or at a linear $e^+e^-$ collider
are outlined.
Finally, the $Q^2$-evolution of the parton content of 
linearly polarized photons, which shows
some remarkable differences, is briefly discussed.
\end{minipage}\\[5mm]

\rule{160mm}{0.4mm}

\end{center}

%
\section{Introduction}
%
The past months have seen a substantial amount of new 
experimental results on
unpolarized deep-inelastic electron-photon scattering from LEP
and LEP2 runs \cite{ref:f2gamma}.
In these measurements of the photon structure function 
$F_2^{\gamma}(x,Q^2)$ the kinematical
coverage in $x$ and $Q^2$ has been considerably extended
as compared to all previous results since PEP and PETRA.
Complementary information on the partonic structure
of photons is provided by increasingly precise photoproduction
measurements at HERA, in particular from (di-)jet production data
\cite{ref:dijet}.
The combination of the available $e^+e^-$ and $ep$ results should
considerably improve our knowledge of the photon structure and
should seriously challenge the various, presently available 
theoretical models.

A similar analysis in longitudinally polarized 
$e^+e^-$ and $ep$ collisions would be desirable.
By measuring the difference between the two independent helicity
combinations of the initial particles, i.e.,
\begin{equation}
\label{eq:xsecdef}
\Delta\sigma = \frac{1}{2}\left[ \sigma(++) - \sigma(+-) \right]
\end{equation}
instead of the sum, as in unpolarized (helicity-averaged) experiments,
one would gain access to the parton structure of longitudinally
(more precisely, circularly) polarized photons, which is completely unmeasured
so far. These densities are defined by
\begin{equation}
\label{eq:pdfdef}
\Delta f^{\gamma}(x,Q^2) = f_+^{\gamma_{+}}(x,Q^2) -  
f_-^{\gamma_{+}}(x,Q^2)\;\;,
\end{equation}
where $f_+^{\gamma_{+}}$ $(f_-^{\gamma_{+}})$ 
denotes the density of a parton $f$ with helicity `+' (`$-$') 
in a photon with helicity `+'. 
The densities $\Delta f^{\gamma}$ contain information different
from that contained in the more familiar unpolarized ones (defined
by taking the sum in (\ref{eq:pdfdef})) and their measurement would
complete our understanding of the partonic structure of photons.

The complete NLO QCD framework for the $Q^2$-evolution of the densities
$\Delta f^{\gamma}(x,Q^2)$ and the calculation of the polarized photon
structure function $g_1^{\gamma}(x,Q^2)$ has become available 
recently with the calculation of the required spin-dependent two-loop
parton-parton \cite{ref:splfct} and photon-parton splitting functions 
\cite{ref:nloletter} and will be reviewed here.
Although such a study seems to be somewhat premature at the first sight
in view of the lack of {\em{any}} experimental 
information on $\Delta f^{\gamma}$ up to now, 
interesting theoretical questions arise when going 
beyond the leading order. Apart from getting
a feeling for the typical size of the NLO corrections, it is moreover
important to analyze the necessity (and feasibility) to introduce a suitable
factorization scheme which overcomes expected problems with perturbative 
instabilities arising in the $\overline{\rm{MS}}$ scheme, in particular for 
large values of $x$ (already known from the unpolarized case, see, 
e.g., \cite{ref:disgamma}).

Furthermore, it is no longer inconceivable to longitudinally polarize
also the proton beam at HERA \cite{ref:polhera}, i.e., to 
run HERA in a polarized collider mode. 
Measurements of spin asymmetries in, e.g., 
the photoproduction of large-$p_T$ (di-)jets 
can then in principle reveal information on the 
$\Delta f^{\gamma}$ through the presence of `resolved' photon 
processes (similar to the already extensively studied case 
with unpolarized beams).
Future polarized linear $e^+e^-$ colliders could provide additional
information on the $\Delta f^{\gamma}$ by measuring the
spin-dependent photon structure function $g_1^{\gamma}(x,Q^2)$ or
spin asymmetries in `resolved' two-photon reactions.
To estimate the feasibility to pin down the so far unknown 
$\Delta f^{\gamma}$ in such experiments, one has to
invoke some theoretical models for $\Delta f^{\gamma}$. 
Moreover, recent progress in calculating spin-dependent 
cross sections up to NLO QCD \cite{ref:nlopola,ref:nlopolb} demonstrates the 
demand for some model distributions even in NLO QCD in order to study 
the impact of the NLO corrections or the remaining scale dependence 
in a consistent manner.

A brief survey of the required theoretical framework 
for the $Q^2$-evolution of the densities $\Delta f^{\gamma}$ in NLO QCD
is given in Sec.~2,
including a discussion of the photonic structure
function $g_1^{\gamma}$, different factorization schemes, and
the so-called `asymptotic' solution for $\Delta f^{\gamma}$. 
Two different scenarios for the $\Delta f^{\gamma}$ 
are presented in Sec.~3 alongside a discussion of theoretical 
constraints on the $\Delta f^{\gamma}$.
In Sec.~4 experimental prospects for measuring the $\Delta f^{\gamma}$
are discussed, with special emphasis on the photoproduction of 
(di-)jets at a polarized HERA. 
Finally, Sec.~5 is devoted to a brief discussion of
the theoretical framework for linearly polarized photons, where some
interesting new features arise.
\vspace*{-0.3cm}
%
\section{$Q^2$-evolution, factorization schemes and all that}
%
Since the photon is a genuine elementary particle, it can directly
interact in hard scattering processes, in addition to its partonic
quark and gluon content $\Delta q^{\gamma}$ and $\Delta g^{\gamma}$,
respectively. Therefore the latter distributions obey the well-known
{\em inhomogeneous} evolution equations schematically given by\footnote{
We follow closely the notation adopted in the unpolarized
case as presented in Refs.~\cite{ref:gr} and \cite{ref:disgamma}.}
\vspace*{-0.1cm}
\begin{equation}
\label{gl1}
\frac{d \Delta q_i^{\gamma}(x,Q^2)}{d \ln Q^2} = \Delta k_i(x,Q^2)+
\left( \Delta P_i \ast \Delta q_i^{\gamma} \right) (x,Q^2) \;\; ,
\end{equation}
where $i$ stands for the flavor non-singlet (NS) quark combinations
or the singlet (S) vector 
$\Delta \vec{q}^{\,\gamma}_{S} \equiv {\Delta \Sigma^{\gamma}
\choose \Delta g^{\gamma}} $, where
$\Delta \Sigma^{\gamma} \equiv \sum_f (\Delta f^{\gamma}+
\Delta \bar{f}^{\gamma} )$ with $f$ running over all active
quark flavors $(f=u,d,s)$. The symbol $\ast$ denotes the usual 
convolution in Bjorken-$x$ space. 
The polarized photon-to-parton and parton-to-parton splitting functions,
$\Delta k_i(x,Q^2)$ and $\Delta P_i(x,Q^2)$, respectively,
in Eq.~(\ref{gl1}) receive the following 1-loop (LO) and 2-loop (NLO) 
contributions:
\vspace*{-0.25cm}
\begin{eqnarray}
\label{gl3}
\nonumber
\Delta k_i(x,Q^2) &=& \frac{\alpha}{2 \pi} \Delta k_i^{(0)}(x)+
\frac{\alpha \alpha_s(Q^2)}{(2\pi)^2} \Delta k_i^{(1)}(x) \\
\Delta P_i(x,Q^2) &=& \frac{\alpha_s(Q^2)}{2 \pi}
\Delta P_i^{(0)}(x) + \left(\frac{\alpha_s(Q^2)}{2 \pi}
\right)^2 \Delta P_i^{(1)}(x) \;\;.
\end{eqnarray}
In the singlet (S) case Eq.~(\ref{gl1})
becomes, of course, a coupled $2\times 2$ matrix equation \cite{ref:nloletter}.
The $\Delta P_{ff'}^{(0,1)}$ can be found in \cite{ref:splfct} and, apart
from obvious NS and S charge factors \cite{ref:nloletter},
the spin-dependent photon-to-parton splitting function
$\Delta k_q^{(0)}$ can be obtained from $\Delta P_{qg}^{(0)}$
by multiplying it with $N_f N_C/T_F$, where $N_C=3$, $T_F=N_f/2$
and $N_f$ being the number of active flavors;
similarly the NLO quantities $\Delta k_q^{(1)}$ and
$\Delta k_g^{(1)}$ correspond to the $C_F T_F$ terms
of $\Delta P_{qg}^{(1)}$ and $\Delta P_{gg}^{(1)}$,
respectively, multiplied by $N_f N_C/T_F$ and are listed in 
\cite{ref:nloletter}\footnote{
Note that $\Delta k_g^{(0)}=0$ due to the missing 
photon-gluon coupling in lowest order.}. 

The evolution equations (\ref{gl1}) 
are most conveniently solved directly in
Mellin-$n$ space, where the solutions can be given analytically.
Taking the $n$th moment of Eq.~(\ref{gl1}), the various
convolutions simply factorize. The required moments of the 
LO and NLO $\Delta k^{(j)}$ and $\Delta P_{ff'}^{(j)}$ $(j=0,1)$
can be found in \cite{ref:nloletter} and \cite{ref:grsv}, respectively,
along with the prescriptions for an analytic continuation in $n$, 
which is required for a numerical Mellin inversion back into $x$ space.
The solution of Eq.~(\ref{gl1}) can be decomposed into a `pointlike' 
(inhomogeneous\footnote{By definition, the pointlike part satisfies
the boundary condition $\Delta q_{i,PL}^{\gamma ,n}(\mu^2)=0$ 
at the input scale $\mu$.}) and a `hadronic' (homogeneous) part, 
i.e.,
\begin{equation}
\label{gl8}
\Delta q_i^{\gamma ,n}(Q^2) = \Delta q^{\gamma ,n}_{i,PL}(Q^2) +
\Delta q^{\gamma ,n}_{i,had}(Q^2)
\end{equation}
($i=$ NS, S) and can be found in \cite{ref:disgamma} (with the obvious 
replacements of all unpolarized quantities by the corresponding
polarized ones, e.g.,  $k_i^{(1)n} \rightarrow \Delta k_i^{(1)n}$).
Having solved the evolution equations (\ref{gl1}) for
$\Delta q_{NS}^{\gamma ,n}(Q^2)$, $\Delta \Sigma^{\gamma ,n}(Q^2)$,
and $\Delta g^{\gamma ,n}(Q^2)$, one finally obtains the desired
$\Delta f^{\gamma ,n}(Q^2)$ $(f=u,\,d,\,s,\,g)$ 
by a straightforward flavor decomposition.

Turning now to spin-dependent deep-inelastic electron-photon
scattering, which can be parametrized in terms of the polarized
structure function $g_1^{\gamma}(x,Q^2)$ (in analogy to the
helicity-averaged case with $F_2^{\gamma}$ and $F_L^{\gamma}$).
In moment-$n$ space the NLO expression for $g_1^{\gamma}$ 
is given by \cite{ref:nloletter} 
(note that $\Delta f^{\gamma}=\Delta \bar{f}^{\gamma}$)
\begin{eqnarray}
\label{gl9}
\nonumber
g_1^{\gamma ,n}(Q^2) &=&  \frac{1}{2} \sum_{f=u,d,s} e_f^2\;
\Bigg\{ 2 \Delta f^{\gamma ,n}(Q^2) 
+ \frac{\alpha_s (Q^2)}{2\pi} \bigg[ 2 \Delta C_q^n 
\Delta f^{\gamma ,n}(Q^2) \\
&& +
\frac{1}{N_f} \Delta C_g^n \Delta g^{\gamma ,n}(Q^2)\bigg] \Bigg\}  
+ \frac{1}{2} N_f N_C \langle e^4\rangle \frac{\alpha}{2\pi} 
\Delta C_{\gamma}^n
\end{eqnarray}
with the usual hadronic spin-dependent Wilson coefficients $\Delta C_q^n$ and
$\Delta C_g^n$, which in the conventional $\overline{\rm{MS}}$ scheme  
can be found, e.g., in \cite{ref:kod,ref:splfct}. 
The photonic coefficient $\Delta C_{\gamma}^n$ can be easily 
derived from $\Delta C_g^n$ and is given in \cite{ref:nloletter}, but 
for the discussions below it might be useful to quote here 
its explicit $x$-space expression
\begin{equation}
\label{gl11}
\Delta C_{\gamma}(x) = 2 \left[(2x-1) \left(\ln \frac{1-x}{x} -1\right)
+2(1-x)\right]\;\;\;.
\end{equation}
The LO expression for $g_1^{\gamma}$ is
entailed in the above formula (\ref{gl9}) by simply dropping
all NLO terms.
Only the contribution of the light flavors has been written out in 
(\ref{gl9}). Heavy quark contributions to $g_1^{\gamma}$ should
be included via the massive polarized direct 
and resolved fusion subprocesses (see, e.g., \cite{ref:gsv}).
These expressions are only available in LO so far.

Let us now turn to the specific features of different factorization
schemes in NLO. It is convenient to introduce a decomposition 
of $g_1^{\gamma ,n}(Q^2)$ analogously to Eq.~(\ref{gl8}): 
\begin{equation}
\label{g1decomp}
g_1^{\gamma ,n}(Q^2) \equiv g_{1,PL}^{\gamma ,n}(Q^2) + 
g_{1,had}^{\gamma ,n}(Q^2)  \; ,
\end{equation}
where $g_{1,PL}^{\gamma ,n}(Q^2)$ is obtained from Eq.~(\ref{gl9})
by taking only $\Delta f^{\gamma ,n}(Q^2)=\Delta f^{\gamma ,n}_{PL}(Q^2)$ 
with $\Delta f^{\gamma ,n}_{PL}(Q^2)$ as defined in (\ref{gl8}).
Conversely, for $g_{1,had}^{\gamma ,n}(Q^2)$ one uses the 
$\Delta f^{\gamma ,n}_{had}(Q^2)$ of (\ref{gl8}), and one obviously has to
omit the $\Delta C_{\gamma}^n$ term in (\ref{gl9}) in this case.

The solutions for $\Delta f^{\gamma ,n}(Q^2)$ ($\Delta f^{\gamma}(x,Q^2)$)
depend on the up to now unspecified hadronic input distributions at the
input scale $Q^2=\mu^2$, i.e., on the boundary conditions for the hadronic
pieces $\Delta f_{had}^{\gamma,n}$ in (\ref{gl8}), which one would 
intuitively relate to some model inspired by vector meson dominance (VMD).
On the other hand, beyond LO both the pointlike as well as the 
hadronic pieces in (\ref{gl8}) depend on the factorization scheme chosen,
and it is a priori not clear in which type of factorization schemes
it actually makes sense to impose a pure VMD hadronic input.
Indeed, in the unpolarized case it was observed that 
\cite{ref:disgamma} the $\ln (1-x)$ term in 
the photonic coefficient function $C_{2,\gamma}(x)$ for $F_2^{\gamma}$,
which becomes negative and divergent for $x\rightarrow 1$, 
drives the pointlike part of $F_2^{\gamma}(x,Q^2)$ in the $\overline{\rm{MS}}$
scheme to large negative values as $x\rightarrow 1$, leading to a strong
difference between the LO and the NLO results for $F_{2,PL}^{\gamma}$ in the
large-$x$ region. As illustrated in Fig.~1, a very similar thing happens 
in the polarized case: here it is the $\ln (1-x)$ term in the polarized 
photonic coefficient function $\Delta C_{\gamma}(x)$ (see Eq.~(\ref{gl11}))
for $g_1^{\gamma}$ that causes large negative values of the pointlike part 
of $g_1^{\gamma}(x,Q^2)$ in the $\overline{\rm{MS}}$ scheme as 
$x\rightarrow 1$, strongly differing from the corresponding LO result 
also shown in Fig.~1. Clearly, the addition of a VMD-inspired hadronic part 
$\Delta f^{\gamma ,n}_{had}(Q^2)$ cannot be sufficient to cure this observed 
instability of $g_{1,PL}^{\gamma}$ in the large-$x$ region since any VMD 
input vanishes as $x\rightarrow 1$. Instead,
as in the unpolarized case, an appropriately adjusted 
non-VMD hadronic NLO input would be required in the $\overline{\rm{MS}}$ 
scheme, substantially differing from the LO one, as the only means of 
avoiding physically not acceptable perturbative instabilities for 
physical quantities like $g_1^{\gamma}(x,Q^2)$.

%
%
\begin{figure}[th]
\begin{center}
\vspace*{-2.5cm}
\epsfig{file=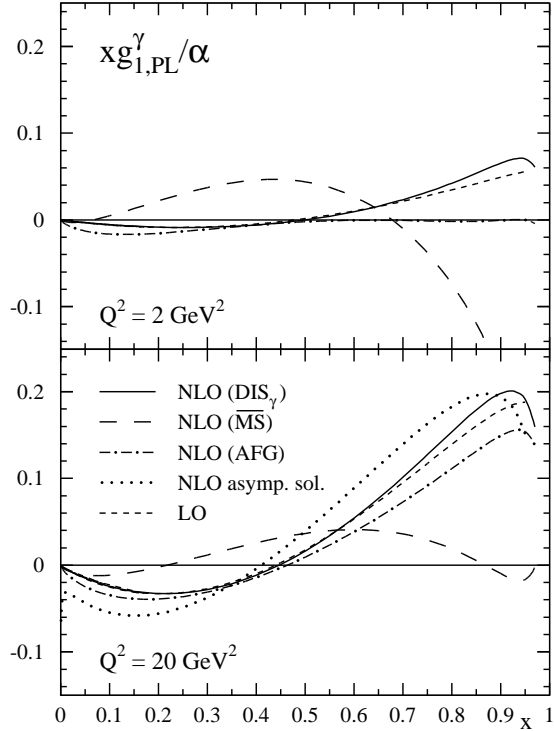,height=13cm}
\vspace*{-1.4cm}
\caption{\sf The `pointlike' part of $x g_1^{\gamma}/\alpha$ (see 
Eq.~(\ref{g1decomp})) in LO and NLO for the $\overline{\rm{MS}}$ and the
$\rm{DIS}_{\gamma}$ factorization schemes. Also shown is the result obtained
when extending the factorization scheme of \cite{ref:aur} to the polarized case
(`AFG', see \cite{ref:nloletter} for details). 
The toy input scale $\mu=1$ GeV, the QCD scale 
parameter $\Lambda=200$ MeV and $N_f=3$ flavors have been used.
For illustration the NLO `asymptotic' solution (see text and 
\cite{ref:nloletter}) is included for $Q^2=20\,{\rm{GeV}}^2$.}
\end{center}
\vspace*{-0.75cm}
\end{figure}
In the unpolarized case the so-called $\rm{DIS}_{\gamma}$ 
scheme \cite{ref:disgamma}
was introduced to avoid such `inconsistencies' by absorbing the photonic
Wilson coefficient for $F_2^{\gamma}$ into the photonic quark distributions.
Analogously, one expects that a similar procedure for the coefficient
$\Delta C_{\gamma}$ for $g_1^{\gamma}$ cures the problem observed 
for $g_{1,PL}^{\gamma}$ in the $\overline{\rm{MS}}$ scheme. This 
redefinition of the polarized photonic quark distributions implies, of course,
also a transformation of the NLO photon-to-parton splitting functions 
$\Delta k_i^{(1)}$ due to the requirement that the physical quantity 
$g_1^{\gamma}$ has to be scheme independent (see \cite{ref:nloletter}
for details). 
The result for $g_{1,PL}^{\gamma}$ after the transformation   
to the $\rm{DIS}_{\gamma}$ scheme is also shown in Fig.~1. The similarity
between the NLO ($\rm{DIS}_{\gamma}$) and the LO curves strongly suggests 
that it is indeed recommendable also in the polarized case to work in the 
$\rm{DIS}_{\gamma}$ scheme.
Moreover, the $\rm{DIS}_{\gamma}$ scheme, also 
eliminates all terms $\sim \ln^2 x$ from the polarized NLO 
$\Delta k_i^{(1)}(x)$, i.e., removes 
the $\overline{\rm{MS}}$ terms leading for $x\rightarrow 0$
(for corresponding observations in the unpolarized case see 
\cite{ref:smallx}).    

The $\Delta f^{\gamma}$ in the $\overline{\rm{MS}}$ 
and the $\rm{DIS}_{\gamma}$ scheme are simply related by \cite{ref:nloletter}
\begin{equation}
\label{gl22}
\Delta f^{\gamma}_{\overline{\rm{MS}}}(x,Q^2) =
\Delta f^{\gamma}_{\rm{DIS}_{\gamma}}(x,Q^2)+\delta \Delta f^{\gamma}(x,Q^2)
\end{equation}
with
\begin{equation}
\label{gl23}
\delta \Delta q^{\gamma}(x,Q^2)=
\delta \Delta \bar{q}^{\gamma}(x,Q^2) = -N_C e_q^2 \frac{\alpha}{4 \pi}
\Delta C_{\gamma}(x)\;,\;\;\;\delta \Delta g^{\gamma}(x,Q^2)=0  \; ,
\end{equation}
where $\Delta C_{\gamma}(x)$ is given in Eq.~(\ref{gl11}).

Let us finish this technical section with a short comment on the so-called
`asymptotic' solution for the $\Delta f^{\gamma}$, which is obtained by 
dropping all terms in the full solution which decrease with increasing 
values of $Q^2$. In this way all dependence on the input scale and the 
boundary conditions is eliminated, and one ends up with the 
unique QCD prediction (see \cite{ref:asy,ref:gr,ref:disgamma} 
for a discussion of the asymptotic solution in the unpolarized case)
\begin{equation}
\label{asympt}
\Delta \vec{q}^{\gamma ,n}_{PL}(Q^2) = \frac{4\pi}{\alpha_s (Q^2)} 
\Delta \vec{a}^n + \Delta \vec{b}^n  \; ,
\end{equation}
where $\Delta \vec{a}^n$ and $\Delta \vec{b}^n$ depend on
the splitting functions (see \cite{ref:nloletter}). 
However, the practical utility of the
asymptotic solution is {\em very limited} since it only applies 
at very large $Q^2$ and $x$: the determinants of the denominators 
in $\Delta \vec{a}^n$ and $\Delta \vec{b}^n$ 
can vanish, causing completely unphysical poles of the 
asymptotic solution which are {\em not} present in the full solution where
subleading ('non-asymptotic') terms regulate such pole terms.
This implies, for instance, that the NLO singlet asymptotic solution
will rise as $\approx x^{-1.57}$ as $x\rightarrow 0$, i.e., the 
asymptotic result for $g_1^{\gamma}$ will {\em not} 
be integrable anymore \cite{ref:nloletter},
whereas for the full solution the first moment of $g_{1,PL}^{\gamma}$
is conserved 
\begin{equation}
\label{eq:g1pl}
\int_0^1 g_{1,PL}^{\gamma}(x,Q^2) dx = 0 \;\;.
\end{equation}
This clearly underlines that the asymptotic solution 
as considered in LO in \cite{ref:xu} can in general 
{\em not} be regarded as a reliable or 
realistic estimate for the polarized photon structure.
%
\section{Available models and theoretical constraints}
%
As already mentioned one has to fully rely
on theoretical models for the polarized photon densities
$\Delta f^{\gamma}$ for the time being.
However, certain theoretical constraints on $\Delta f^{\gamma}$
might have to be taken into account when constructing such models:
\begin{itemize}
\item{\bf `Positivity':} Positivity of the helicity dependent
cross sections on the r.h.s.\ of Eq.~(\ref{eq:xsecdef}) demands that
$|\Delta \sigma| \le \sigma$, which can be directly translated into
a useful constraint on the 
densities\footnote{Strictly speaking positivity applies
only to physical quantities like cross sections and not to
parton densities beyond the LO where they become scheme-dependent 
(`unphysical') objects. Of course, (\ref{eq:pos}) still serves as a
reasonable `starting point' for the NLO densities as (\ref{eq:pos})
is preserved by the NLO evolution kernels.}  
\begin{equation}
\label{eq:pos}
|\Delta f^{\gamma}(x,Q^2)| \le f^{\gamma}(x,Q^2)\;\;.
\end{equation}
\item{\bf `Current conservation':} In \cite{ref:currentcons} it was shown 
that the first moment of $g_1^{\gamma}$ vanishes irrespective of $Q^2$.
This result is non-perturbative: it holds to all orders in perturbation
theory and at every twist provided that the fermions in the theory 
have non-vanishing mass \cite{ref:currentcons}.
Due to Eq.~(\ref{eq:g1pl}) this sum rule can be realized in LO and NLO
($\overline{\mathrm{MS}}$ or $\mathrm{DIS}_{\gamma}$) by demanding
\begin{equation}
\label{eq:curcons}
\Delta q_{had}^{\gamma,n=1}=0\;\;,
\end{equation}
i.e., a vanishing first moment of the photonic quark densities at the input
scale (the gluon input is not constrained since $\Delta C_g^{n=1}=0$ in
(\ref{gl9})).
\end{itemize}

To obtain a realistic estimate for the 
theoretical uncertainties in the polarized photon structure functions 
coming from the unknown hadronic input, 
we consider two very different scenarios 
in LO \cite{ref:gv,ref:gsv} and NLO ($\mathrm{DIS}_{\gamma}$)
\cite{ref:nloletter} based on the positivity bound (\ref{eq:pos}):
for the first (`maximal scenario') we saturate (\ref{eq:pos})
using the phenomenologically successful unpolarized GRV photon densities
\cite{ref:grvphot} 
\begin{equation}
\label{eq:maxsat} 
\Delta f^{\gamma}_{had}(x,\mu^2) = f_{had}^{\gamma}(x,\mu^2)\;\;,
\end{equation}
whereas the other extreme input (`minimal scenario') is defined by
\begin{equation}
\label{eq:minsat}
\Delta f^{\gamma}_{had}(x,\mu^2) = 0
\end{equation}
with $\mu=\mu_{LO,NLO}\simeq 0.6\,{\rm{GeV}}$ \cite{ref:grvphot}.
Of the two extreme hadronic inputs only the `minimal' one 
(Eq.~(\ref{eq:minsat})) satisfies (\ref{eq:curcons}).
However, we are interested only in the region of, say,
$x>0.01$ here, such that for the `maximal' scenario (\ref{eq:maxsat}) the 
constraint (\ref{eq:curcons}) could well be implemented by contributions 
from smaller $x$ which do not affect the evolutions at larger $x$. 
In addition the sum rule is not expected to hold in massless QCD.
Rather than artificially enforcing the vanishing of the first moment 
of the $\Delta q_{had}^{\gamma} (x,\mu^2)$ in the `maximal' scenario, 
we therefore stick to the two extreme scenarios as introduced above.

%
%
\begin{figure}[bth]
\vspace*{-1.4cm}
\begin{center}
\epsfig{file=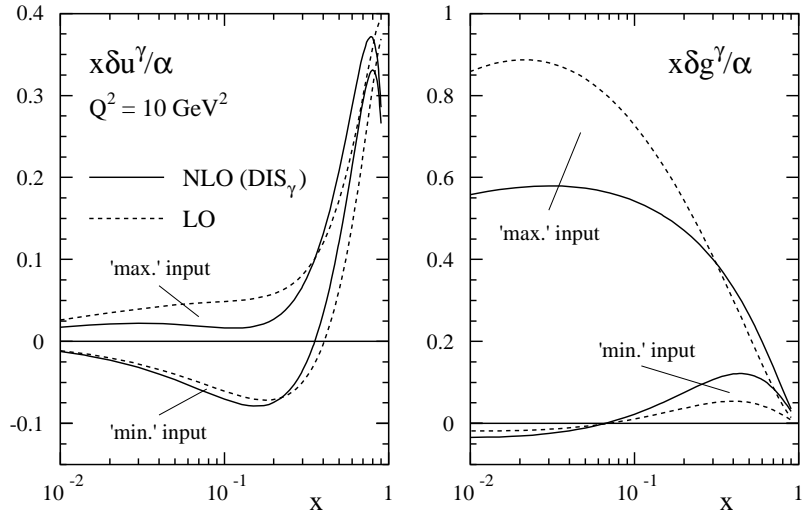, width=12cm}
\vspace*{-1.1cm}
\caption{\sf The LO and NLO ($\rm{DIS}_{\gamma}$) polarized 
photonic parton densities according to the `maximal' and `minimal' inputs
of Eqs.~(\ref{eq:maxsat}) and (\ref{eq:minsat}), respectively,
evolved to $Q^2=10\;\mathrm{GeV}^2$.}
\end{center}
\vspace*{-0.5cm}
\end{figure}
In Fig.~2 we compare our LO and NLO ($\rm{DIS}_{\gamma}$)  
distributions $x \Delta u^{\gamma}/\alpha$, $x \Delta g^{\gamma}/\alpha$ 
for the two extreme scenarios at $Q^2=10\,{\rm{GeV}}^2$. 
These two extreme sets should be useful and sufficient in studies
of the prospects of future spin experiments.
%
\section{Experimental prospects}
%
Due to the lack of space, we concentrate here mainly on jet photoproduction
at a polarized HERA, which seems to be the most promising tool to gain
some information about the $\Delta f^{\gamma}$ (similar unpolarized
measurements at HERA have already successfully reduced our ignorance 
of the helicity-averaged densities $f^{\gamma}$). 
A detailed study of the physics case of the polarized collider mode 
option for HERA can be found in \cite{ref:polhera}.

Schematically, any photoproduction cross section $(\Delta) \sigma$ is a
sum of a so-called `direct' and a `resolved' photon contribution:
\begin{equation}
\label{eq:dirres}
(\Delta) \sigma = (\Delta)f^p \ast (\Delta) \hat{\sigma}_{\gamma f} +
(\Delta)f^p \ast (\Delta) f'^{\gamma} \ast (\Delta)
\hat{\sigma}_{ff'}\;\;.
\end{equation}
Apart from the $\Delta f^{\gamma}$, the polarized parton densities  
$\Delta f^p$ of the proton enter in (\ref{eq:dirres}) as well, 
which complicates
a determination of the $\Delta f^{\gamma}$ since the $\Delta f^p$,
especially $\Delta g^p$, are not
very well constrained by presently available DIS data 
(see, e.g., \cite{ref:grsv}).
However, at the time polarized HERA could be operational, new information
on the $\Delta f^p$, in particular on $\Delta g^p$, will be available
from the upcoming experiments COMPASS at CERN and BNL-RHIC.

%
%
\begin{figure}[th]
\vspace*{-1.4cm}
\begin{center}
\epsfig{file=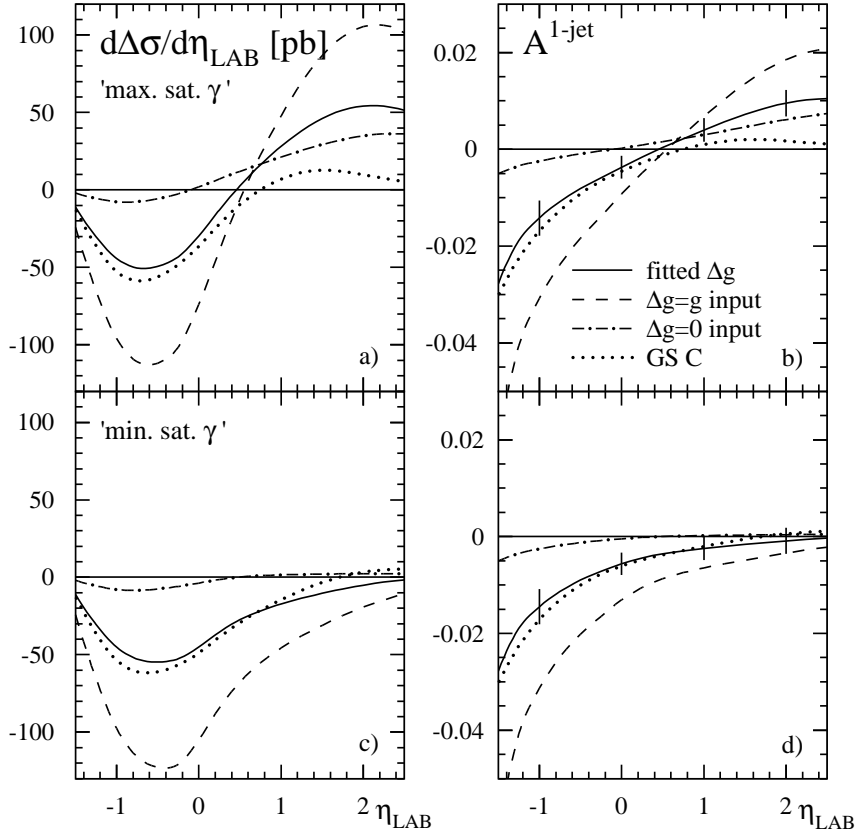, width=12.5cm}
\vspace*{-1.0cm}
\caption{\sf {\bf{a:}} $\eta_{LAB}$ dependence of the 
polarized single-jet inclusive
photoproduction cross section at HERA, integrated over
$p_T>8\,\mathrm{GeV}$ (see \cite{ref:svhera} for details).
The resolved contribution to the cross section has been
calculated with the `maximally' saturated set of polarized photonic
densities. {\bf{b:}} Asymmetry corresponding to {\bf{a}}. 
Also shown are the expected statistical errors, see \cite{ref:svhera}. 
{\bf{c,d:}} Same as {\bf{a}}, {\bf{b}}, but for
the `minimal' scenario of the $\Delta f^{\gamma}$.}
\end{center}
\vspace*{-0.8cm}
\end{figure}
Studying laboratory frame rapidity distributions at HERA\footnote{As
conventional for HERA, $\eta_{LAB}$ is defined to be positive in the proton
forward direction.} is a particularly suited way of `separating' off the
direct from the resolved contributions in (\ref{eq:dirres}) in a
single-inclusive measurement of jets or hadrons \cite{ref:svhera}: for negative
$\eta_{LAB}$ the main contribution is expected to come from the region
of $x_{\gamma}\rightarrow 1$ and thus mostly from the direct part. The
situation is reversed at positive $\eta_{LAB}$, where one should become 
sensitive to the unknown $\Delta f^{\gamma}$.
To investigate this conjecture, Fig.~3 shows our results for the
polarized single-inclusive jet cross section and its asymmetry 
$A^{1-jet}\equiv \Delta \sigma/\sigma$, which is the experimentally
relevant quantity, vs.\ $\eta_{LAB}$ for four different sets
of the $\Delta f^p$ (for more details see \cite{ref:svhera}).
For Figs.~3a,b we have used the `maximally' saturated set of 
$\Delta f^{\gamma}$, 
whereas Figs.~3c,d correspond to the `minimally' saturated one.
A comparison of these results shows indeed a sensitivity to the 
different $\Delta f^{\gamma}$ only in the 
forward direction for $\eta_{LAB}\ge 1$,
where the resolved contribution dominates. Fig.~3 also includes the
expected statistical errors for such a measurement at HERA 
\cite{ref:svhera} assuming an 
integrated luminosity ${\cal{L}}=100\;\mathrm{pb}^{-1}$.
Clearly, one can really learn something about the completely unknown polarized
photon structure, {\em provided} the proton densities $\Delta f^p$ are pinned
down more precisely in the future.
Similar results can be obtained from single-inclusive charged hadron 
production \cite{ref:svhera,ref:hera2}. It should be also 
noted that the shown LO asymmetries in
Fig.~3 are rather stable under variations of the factorization 
scale\footnote{Very recently, the complete NLO corrections were 
calculated \cite{ref:nlopola},
but no phenomenological studies have been performed yet.}
in contrast to the polarized LO cross sections $\Delta \sigma$.

%
%
\begin{figure}[th]
\vspace*{-1.5cm}
\begin{center}
\epsfig{file=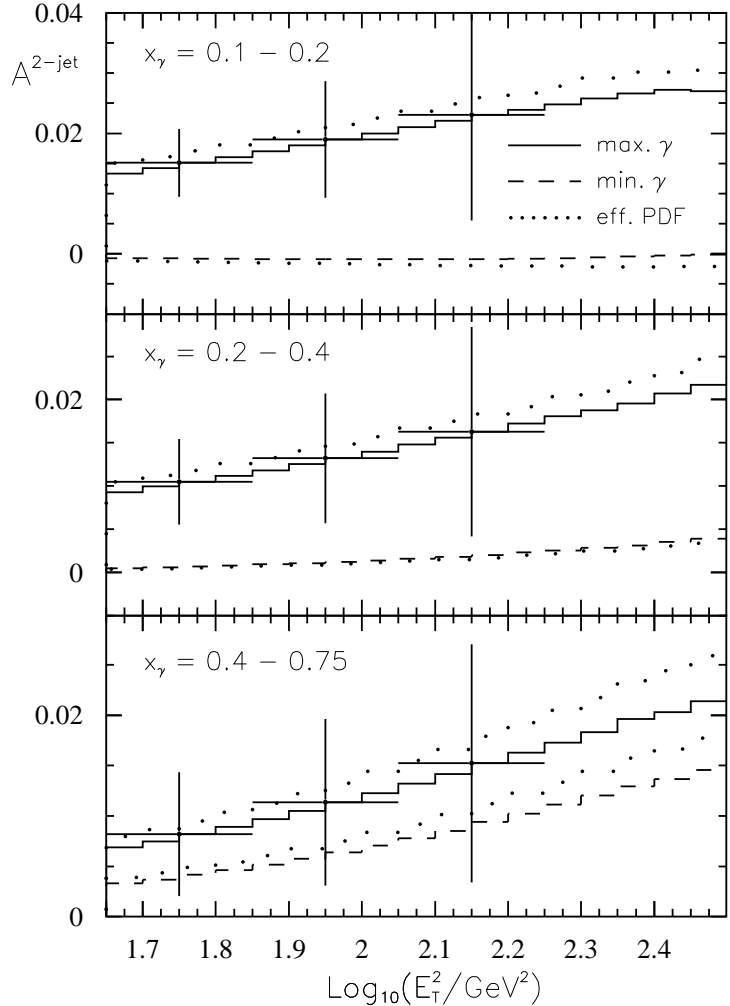, width=11.0cm}
\vspace*{-0.8cm}
\caption{\sf The LO di-jet spin asymmetry in different $x_{\gamma}$
bins as a function of the jet transverse energy $E_T^{jet}$ for
the two extreme set of $\Delta f^{\gamma}$. 
Cuts on the jet rapidities are as in the recent unpolarized
measurement by H1 \cite{ref:h1dijet}. The dotted lines
correspond to the approximation as described in the text.  }
\end{center}
\vspace*{-0.8cm}
\end{figure}
Even more promising is dijet production. The important point here is that
such a measurement allows for fully reconstructing the kinematics
of the underlying hard process, i.e., experimentally determining the
momentum fraction $x_{\gamma}$ for each event. Thus it becomes 
possible to experimentally select a resolved sample by taking only
events with, say, $x_{\gamma}\le 0.75$. Again with 
${\cal{L}}=100\;\mathrm{pb}^{-1}$ a decent measurement should be
possible \cite{ref:svhera,ref:hera2}. In 
\cite{ref:hera2} we have furthermore shown that the LO QCD parton
level calculations do not differ too much from MC results including QCD 
radiation, hadronization, etc.

Fig.~4 shows the LO 2-jet asymmetry in three different $x_{\gamma}$-bins
for the two extreme $\Delta f^{\gamma}$ sets. For the $\Delta f^p$ the 
GRSV LO `standard' distributions \cite{ref:grsv} have been used (of course,
as for Fig.~3, there is a potential ambiguity due to our
present ignorance of $\Delta g^p$, which has to be reduced before any
information on the $\Delta f^{\gamma}$ can be obtained).
Based on a rather old idea \cite{ref:effpdf}, the various different resolved
subprocess cross sections and parton combinations entering the
analysis can be approximated by an effective cross section which depends
only on a certain combination of the proton and photon densities
\begin{equation}
\label{eq:effpdf}
(\Delta) f_{eff}^{\gamma,p} = 
\sum_q \left[ (\Delta)q + (\Delta) \bar{q}\right]
+ (\Delta)a\, (\Delta)g\;\;,
\end{equation}
where $a=9/4$ \cite{ref:effpdf} and $\Delta a=11/4$ \cite{ref:prep}.
The result of this rather accurate approximation is also shown in
Fig.~4 (dotted lines). The unpolarized 
combination $f^{\gamma}_{eff}$ was recently measured by H1 at HERA
using this method \cite{ref:h1dijet}, and the prospects of 
unfolding $\Delta f^{\gamma}_{eff}$
in a similar way are currently under investigation \cite{ref:prep}.

It should be mentioned here that heavy flavor production at a polarized
HERA is not suited to determine either $\Delta g^p$ 
or the $\Delta f^{\gamma}$.
A similar remark applies to prompt photon production. In both cases
the statistical accuracy would be far too limited \cite{ref:hera2}.
The helicity transfer $\Lambda$-baryon photoproduction process 
$\vec{e}p\rightarrow \vec{\Lambda} X$, which can be studied 
{\em without} polarized protons at HERA, is, unfortunately, mainly sensitive
to the polarized $\Lambda$ fragmentation functions $\Delta D_f^{\Lambda}$ 
and only to a much lesser extent to the $\Delta f^{\gamma}$
\cite{ref:lambda}.
However, such a measurement can shed some light on the also poorly 
understood spin-dependent fragmentation functions, which is an issue
interesting in its own.
 
Finally, some brief remarks about the prospects of a future polarized
linear $e^+e^-$
collider. This would be a unique place to study for the first time the 
structure function $g_1^{\gamma}$ in spin-dependent deep-inelastic
electron-photon scattering. Moreover, di-jets, etc.\ can be studied
in single and double-resolved photon processes. However, compared to
similar $ep$ cross sections, which are ${\cal{O}}(\alpha^2\alpha_s)$,
${\cal{O}}(\alpha^4)$ $e^+e^-$ reactions are further suppressed and
much higher luminosities are required to keep the statistical errors
small enough to discriminate between different $\Delta f^{\gamma}$
scenarios.
A possible improvement would be the use of polarized backscattered laser 
photons \cite{ref:laser}, i.e., a dedicated polarized
$\gamma\gamma$ collider.
The spectrum of backscattered laser photons can lead to considerably larger
cross sections as compared to the `usual' equivalent photon
approximation and detailed quantitative analyses are currently under way.

%
\section{Linearly polarized photons}
%
Finally, let us turn to the parton content of linearly polarized
photons. There are {\em no} quark densities for this kind of
polarization \cite{ref:linpol} and the linearly polarized gluon
density of the photon is defined by\footnote{It should be noted
that the structure function of linearly polarized photons is 
sometimes denoted also by $F_3^{\gamma}$ in the literature \cite{ref:f3gamma}.}
\begin{equation}
\label{eq:lindef}
\Delta_L g^{\gamma} \equiv g_{\hat{x}}^{\gamma_{\hat{x}}} -
                           g_{\hat{y}}^{\gamma_{\hat{x}}}\;\;,
\end{equation}
where $\hat{x}$ $(\hat{y})$ denotes 
linear polarization along the $x$ ($y$) axis.
The $\Delta_L g^{\gamma}$ obeys a very simple, non-singlet type,
inhomogeneous evolution equation
\begin{equation}
\label{eq:linevol}
\frac{d \Delta_L g^{\gamma}}{d\ln Q^2} = \Delta_L k_g + 
\Delta_L P_{gg} \ast \Delta_L g^{\gamma}\;\;,
\end{equation}
where the relevant linearly polarized splitting functions
$\Delta_L k_g$ and $\Delta_L P_{gg}$ are taken to have a
perturbative expansion as in (\ref{gl3}). $\Delta_L P_{gg}$ was
recently calculated in NLO and can be found in \cite{ref:cracow}.
The $C_F T_F$ part of $\Delta_L P_{gg}^{(1)}$ also determines 
the first non-vanishing,
lowest order  photon-gluon splitting function $\Delta_L k_g^{(1)}$ 
along the same lines as described below Eq.~(\ref{gl3}).

When solving (\ref{eq:linevol}), this ${\cal{O}}(\alpha \alpha_s)$
$\Delta_L k_g^{(1)}$ has to be combined with the {\em lowest order}
${\cal{O}}(\alpha_s)$ gluon-gluon splitting function
$\Delta_L P_{gg}^{(0)}$.
The bottom line is that the partonic structure
of linearly polarized photons in LO QCD 
is {\em{not}} of ${\cal{O}}(\alpha/\alpha_s)$
anymore, as in the case of unpolarized or longitudinally polarized photons 
(see, e.g., (\ref{asympt})), but ${\cal{O}}(\alpha)$, due to the lack of a
${\cal{O}}(\alpha)$ quark `driving term' $\sim (\Delta) k_q^{(0)}$ 
in (\ref{eq:linevol}).
Both terms on r.h.s.\ of Eq.~(\ref{eq:linevol}) are
then ${\cal{O}}(\alpha \alpha_s)$, and a consistent NLO evolution of
$\Delta_L g^{\gamma}$ would require the knowledge of $\Delta_L k_g^{(2)}$.
In addition, all resolved processes are 
$\alpha_s$-suppressed compared to the direct photon
contribution, i.e., are formally NLO effects.
More details will be presented in \cite{ref:linpaper}.
%
\section*{Acknowledgements}
%
I am grateful to Werner Vogelsang for his collaboration on
all topics presented here and to the organizers of the workshop 
for a pleasant and stimulating meeting.
%


\begin{thebibliography}{99}
%
\bibitem{ref:f2gamma} See, for example: M.\ Kienzle, these proceedings;\\
B.\ Surrow, these proceedings;\\
I.\ Tyapkin, these proceedings.
%
\bibitem{ref:dijet} See, for example, Y.\ Yamazaki, these proceedings.
%
\bibitem{ref:splfct}  R.\ Mertig and W.L.\ van Neerven, 
Z. Phys. {\bf{C70}} (1996) 637;\\
W.\ Vogelsang, Phys. Rev. {\bf{D54}} (1996) 2023; 
Nucl. Phys. {\bf{B475}} (1996) 47.
%
\bibitem{ref:nloletter} M.\ Stratmann and W.\ Vogelsang, 
Phys. Lett. {\bf B386} (1996) 370.
%
\bibitem{ref:disgamma} M.\ Gl\"{u}ck, E.\ Reya, and A.\ Vogt,
Phys. Rev. {\bf{D45}} (1992) 3986.
%
\bibitem{ref:polhera} Proceedings of the workshop `Physics with
Polarized Protons at HERA', DESY, 1997, eds.\ A.\ De Roeck and
T.\ Gehrmann, DESY-PROCEEDINGS-1998-01.
%
\bibitem{ref:nlopola} D.\ de Florian, S.\ Frixione, A.\ Signer, and
W.\ Vogelsang, {\tt hep-ph/9808262}.
%
\bibitem{ref:nlopolb} I.\ Bojak and M.\ Stratmann, 
Phys. Lett. {\bf{B433}} (1998) 411; {\tt hep-ph/9807405}.
%
\bibitem{ref:gr} M.\ Gl\"{u}ck and E.\ Reya, Phys. Rev. {\bf{D28}} (1983)
2749.
%
\bibitem{ref:grsv} M.\ Gl\"{u}ck, E.\ Reya, M.\ Stratmann, and
W.\ Vogelsang, Phys. Rev. {\bf{D53}} (1996) 4775. 
%
\bibitem{ref:kod} J.\ Kodaira, S.\ Matsuda, T.\ Muta, K.\ Sasaki, and 
T.\ Uematsu, Phys. Rev. {\bf D20} (1979) 627.
%
\bibitem{ref:gsv} M.\ Gl\"{u}ck, M.\ Stratmann and W.\ Vogelsang,
Phys. Lett. {\bf{B337}} (1994) 373.
%
\bibitem{ref:aur} P.\ Aurenche, M.\ Fontannaz, and J.Ph.\ Guillet,
Z. Phys. {\bf C64} (1994) 621.
%
\bibitem{ref:smallx} A.\ Vogt, in proc.\ of the 
workshop on `Two-Photon Physics
at LEP and HERA', Lund, 1994, eds.\ G.\ Jarlskog and L.\ J\"{o}nsson,
p.141;\\
For the timelike situation (photon fragmentation functions) 
see also M.\ Gl\"{u}ck, E.\ Reya and A.\ Vogt, Phys. Rev. {\bf D48} (1993) 116.
%
\bibitem{ref:asy} E.\ Witten, Nucl. Phys. {\bf B120} (1977) 189;\\
W.A.\ Bardeen and A.J.\ Buras, Phys. Rev. {\bf D20} (1979) 166; 
{\bf D21} (1980) 2041(E).
%
\bibitem{ref:xu} A.C.\ Irving and D.B.\ Newland, Z. Phys. {\bf C6} (1980)
27;\\
J.A.\ Hassan and D.J.\ Pilling, Nucl. Phys. {\bf B187} (1981) 563;\\
Zai-xin Xu, Phys. Rev. {\bf D30} (1984) 1440.
%
\bibitem{ref:currentcons} S.D.\ Bass, Int. J. Mod. Phys. {\bf A7}
(1992) 6039;\\
S.\ Narison, G.M.\ Shore, and G.\ Veneziano, Nucl. Phys. {\bf{B391}}
(1993) 69;\\
S.D.\ Bass, S.J.\ Brodsky, and I.\ Schmidt, {\tt hep-ph/9805316}.
%
\bibitem{ref:gv} M.\ Gl\"{u}ck and W.\ Vogelsang, Z. Phys. {\bf C55}
(1992) 353; {\bf C57} (1993) 309.
%
\bibitem{ref:grvphot} M.\ Gl\"{u}ck, E.\ Reya, and A.\ Vogt,
Phys. Rev. {\bf{D46}} (1992) 1973.
%
\bibitem{ref:svhera} M.\ Stratmann and W.\ Vogelsang, Z. Phys. {\bf C74}
(1997) 641; in proc.\ of the 1995/96 workshop on `Future Physics at HERA',
DESY, Hamburg, eds.\ G.\ Ingelman et al., p.\ 815.
%
\bibitem{ref:hera2} J.M.\ Butterworth, N.\ Goodman, M.\ Stratmann, and
W.\ Vogelsang, in proc. of the workshop on `Physics with Polarized
Protons at HERA', DESY, Hamburg, 1997, eds.\ A.\ De Roeck and T.\ Gehrmann, 
DESY-PROCEEDINGS-1998-01, p.\ 120.
%
\bibitem{ref:h1dijet} C.\ Adloff et al., H1 collab., Eur. Phys. J. {\bf C1}
(1998) 97.
%
\bibitem{ref:effpdf} B.L.\ Combridge and C.J.\ Maxwell, 
Nucl. Phys. {\bf B239} (1984) 429.
%
\bibitem{ref:prep} A.\ De Roeck, G.\ R\"{a}del, M.\ Stratmann, and
W.\ Vogelsang, in preparation.
%
\bibitem{ref:lambda} D.\ de Florian, M.\ Stratmann, and W.\ Vogelsang,
in proc. of the workshop on `Physics with Polarized
Protons at HERA', DESY, Hamburg, 1997, eds.\ A.\ De Roeck and T.\ Gehrmann, 
DESY-PROCEEDINGS-1998-01, p.\ 140.
%
\bibitem{ref:laser} I.F.\ Ginzburg et al., Nucl. Inst. and Meth. {\bf 205}
(1983) 47; {\bf 219} (1984) 5;\\
D.L.\ Borden, D.A. Bauer, and D.O. Caldwell, SLAC-PUB-5715, 1992 
(unpublished); Phys. Rev. {\bf D48} (1993) 4018.
%
\bibitem{ref:linpol} F.\ Delduc, M.\ Gourdin, and E.G.\ Oudrhiri-Safiani,
Nucl. Phys. {\bf B174} (1980) 157;\\
R.L.\ Jaffe and A.\ Manohar, Phys. Lett. {\bf B223} (1989) 218;\\
X.\ Artru and M.\ Mekhfi, Z. Phys. {\bf C45} (1990) 669.
%
\bibitem{ref:f3gamma} 
A.\ Manohar, Phys. Lett. {\bf B219} (1989) 357;\\
B.\ Ermolaev, R.\ Kirschner, and L.\ Szymanowski, {\tt hep-ph/9806439}
(to appear in Eur. Phy. J. {\bf C}).
%
\bibitem{ref:cracow} W.\ Vogelsang, in proc.\ of the conference on
`Spin Effects in Particle Physics and Tempus Workshop', Cracow, 1998,
Acta Phys. Pol. {\bf B29} (1998) 1189. 
%
\bibitem{ref:linpaper} M.\ Stratmann and W.\ Vogelsang, in preparation.
%
\end{thebibliography}
\end{document}